\documentclass[prd,reprint,showpacs,showkeys]{revtex4-1}
\usepackage{amsfonts}
\usepackage{amssymb}
\usepackage{amsmath}
\usepackage{graphicx}
\usepackage[font={footnotesize,it}]{caption}

\begin{document}

\title{Einstein-Maxwell gravity coupled to a scalar field in 2+1-dimensions}
\author{S. Habib Mazharimousavi}
\email{habib.mazhari@emu.edu.tr}
\author{M. Halilsoy}
\email{mustafa.halilsoy@emu.edu.tr}
\affiliation{Department of Physics, Eastern Mediterranean University, Gazima\~{g}usa,
north Cyprus, Mersin 10, Turkey. }
\date{\today }

\begin{abstract}
We consider Einstein-Maxwell-self-interacting scalar field theory described
by a potential $V\left( \phi \right) $ in $2+1-$dimensions. The
self-interaction potential is chosen to be a highly non-linear
double-Liouville type. Exact solutions, including chargeless black holes and
singularity-free non-black hole solutions are obtained in this model.
\end{abstract}

\maketitle

\section{Introduction}

Minimally coupled pure scalar field solutions in curved $2+1-$dimensions is
severely restricted to admit a variety of solutions \cite{1}. This is in
contrast to higher dimensions in which generation of scalar field solutions
follow from known vacuum Einstein solutions. The fact that there is no
vacuum solution in $2+1-$dimensions makes this method inapplicable. The
strategy therefore is followed by adding sources such as cosmological
constant \cite{2,3}, electromagnetic (both linear and non-linear) field \cite%
{4,5,6} and non-minimally coupled scalar fields to couple with such sources 
\cite{7,8,9}. Self-interacting scalar fields as a potential term in the
Lagrangian is also an alternative method to investigate the role / effects
of scalar fields \cite{10,11,12,13,14,15}. A well-known class of
self-interacting scalar fields for instance is given by global monopole \cite%
{16} which arises as a result of spontaneous symmetry breaking. The idea is
to search for black holes with scalar hairs in analogy with the
electromagnetic fields. In this regard scalar fields alone in higher
dimensions ($D>3$) creates mostly naked singularities and very rarely black
holes. The singular solution by Janis, Newman and Winicour \cite{17} is a
prototype in this regard in $3+1-$dimensions. The solution by
Bocharova-Bronnikov-Melnikov-Bekenstein (BBMB) provides a black hole
solution in the same dimensionality \cite{18,19}. The massive scalar field
in three dimensions has been also considered in \cite{20,21,22}.

In this study we consider minimally coupled scalar, Maxwell fields and a
self-interaction potential $V\left( \phi \right) $ for the scalar field. Our
choice for the potential is in the form of a double-Liouville potential \cite%
{23} of the form $V\left( \phi \right) =\lambda _{1}e^{\alpha _{1}\phi
}+\lambda _{2}e^{\alpha _{2}\phi }$ in which $\lambda _{i}$ and $\alpha _{i}$
are constant parameters. It should be added that this form of potential is
not reducible to a single-Liouville potential $V\left( \phi \right) =\lambda
e^{\alpha \phi }$ with constant $\lambda $ and $\alpha $. The occurrence of
extra parameters doesn't create redundancy in the problem and as a matter of
fact it renders the solution possible. Certain solutions may arise in which
some parameters are dispensable. As it will be observed the Liouville-type
potential is too strong and creates singularity at the origin. We show that
black hole solutions can also be obtained along with the non-black hole
solutions in such a model. This happens as a result of tuning our free
parameters. With zero electric charge, for instance, the Einstein-Scalar
system gives rise to a black hole solution with constant Hawking
temperature. More general solutions can be obtained as a result of reduction
of the system of equations into a Riccati type. We find as an example
chargeless, singularity free solution that at infinity becomes conformal to
the BTZ spacetime.

Organization of the paper goes as follows. In Section II we derive the field
equations of our model. Exact black hole solutions are represented with zero
electric charge in Section III. Non-black hole solutions, both singular and
non-singular are given in Section IV. The paper ends with Conclusion in
Section V.

\section{Einstein-Maxwell gravity coupled to a scalar field}

The action of Einstein-Maxwell gravity coupled minimally to a scalar field $%
\phi $ is given by ($16\pi G=c=1$)%
\begin{equation}
S=\int d^{3}x\sqrt{-g}\left( \mathcal{R}-2\partial _{\mu }\phi \partial
^{\mu }\phi -F^{2}-V\left( \phi \right) \right) 
\end{equation}%
in which $F=F_{\mu \nu }F^{\mu \nu }$ stands for the Maxwell invariant and $%
V\left( \phi \right) $ is to be chosen (We would like to note that our
formalism follows the higher dimensional version of \cite{24,25}). The Field
equations are given by applying the variational principles,%
\begin{equation}
d\left( ^{\star }\mathbf{F}\right) =0,
\end{equation}%
\begin{equation}
\square \phi =\frac{1}{4}\frac{dV}{d\phi },
\end{equation}%
and%
\begin{equation}
G_{\mu }^{\nu }=T_{\mu }^{\nu }
\end{equation}%
in which 
\begin{multline}
T_{\mu }^{\nu }=2\left( F_{\mu \lambda }F^{\nu \lambda }-\frac{1}{4}%
F_{\alpha \beta }F^{\alpha \beta }\delta _{\mu }^{\nu }\right) + \\
2\left( \partial _{\mu }\phi \partial ^{\nu }\phi -\frac{1}{2}\partial
_{\lambda }\phi \partial ^{\lambda }\phi \delta _{\mu }^{\nu }\right) -\frac{%
1}{2}V\left( \phi \right) \delta _{\mu }^{\nu },
\end{multline}%
and $\square $ stands for the covariant Laplacian. The spacetime under study
is static and circularly symmetric with a line element of the form%
\begin{equation}
ds^{2}=-U\left( r\right) dt^{2}+\frac{dr^{2}}{U\left( r\right) }+R^{2}\left(
r\right) d\theta ^{2}
\end{equation}%
in which $U\left( r\right) $ and $R\left( r\right) $ are the metric
functions to be found. The electric field ansatz $2-$form is 
\begin{equation}
\mathbf{F}=\frac{q}{R}dt\wedge dr
\end{equation}%
in which $q$ is the electric charge. Considering this ansatz, we find the
following field equations,%
\begin{equation}
\left( RU\phi ^{\prime }\right) ^{\prime }=\frac{R}{4}\frac{dV}{d\phi },
\end{equation}%
\begin{equation}
\frac{R^{\prime \prime }}{R}=-2\phi ^{\prime 2},
\end{equation}%
\begin{equation}
\left( RU\right) ^{\prime \prime }=-3RV-\frac{2q^{2}}{R}
\end{equation}%
and%
\begin{equation}
\left( UR^{\prime }\right) ^{\prime }=-\frac{2q^{2}}{R}-RV.
\end{equation}

\section{An exact black hole solution}

In this chapter we give an exact solution with the double Liouville potential%
\begin{equation}
V\left( \phi \right) =\lambda _{1}e^{\alpha _{1}\phi }+\lambda _{2}e^{\alpha
_{2}\phi }
\end{equation}%
in which $\lambda _{i}$ and $\alpha _{i}$ are some constants. Let us add
that the choice of double-potential term can't be reduced through
reparametrization into a single-Liouville potential. The advantage of
employing more parameters will be clear subsequently. Our ansatz for $R$ is%
\begin{equation}
R=e^{A\phi }
\end{equation}%
in which $A$ is a constant to be found. Plugging these into the field
equations yields the following solution for the scalar field 
\begin{equation}
\phi =\frac{A\ln \left( 1+\frac{r}{r_{0}}\right) }{A^{2}+2}
\end{equation}%
in which $r_{0}$ is an integration constant. With $\alpha _{1}=-\frac{4}{A}$
and $\alpha _{2}=-2A$ one also finds%
\begin{multline}
U\left( r\right) =C_{1}\left( 1+\frac{r}{r_{0}}\right) ^{\frac{2}{A^{2}+2}}-%
\frac{\lambda _{1}r_{0}^{2}\left( A^{2}+2\right) ^{2}}{2A^{2}\left(
A^{2}-1\right) }\left( 1+\frac{r}{r_{0}}\right) ^{\frac{2A^{2}}{A^{2}+2}} \\
-\frac{\left( \lambda _{2}+2q^{2}\right) r_{0}^{2}\left( A^{2}+2\right) ^{2}%
}{2A^{2}}\left( 1+\frac{r}{r_{0}}\right) ^{\frac{4}{A^{2}+2}}
\end{multline}%
in which $C_{1}$ is another integration constant and the parameter $A$ is
given by%
\begin{equation}
A^{2}=2\left( 1+\frac{2q^{2}}{\lambda _{2}}\right) .
\end{equation}%
We note that the electric charge $q$ and $\lambda _{2}$ must be chosen in
such a way that $1+\frac{2q^{2}}{\lambda _{2}}>0$ and $1+\frac{2q^{2}}{%
\lambda _{2}}\neq \frac{1}{2}.$ This guarantees that $A^{2}\neq 0,1$ and
remains positive. The form of the electric field in its closed form reads%
\begin{equation}
E=qe^{-A\phi }=q\left( \frac{r_{0}}{r+r_{0}}\right) ^{\frac{A^{2}}{A^{2}+2}}.
\end{equation}%
It should be added that by introducing $\rho =1+\frac{r}{r_{0}}$ and $\tau
=r_{0}t$ the line element becomes%
\begin{equation}
ds^{2}=-f\left( \rho \right) d\tau ^{2}+\frac{d\rho ^{2}}{f\left( \rho
\right) }+\rho ^{\frac{2A^{2}}{A^{2}+2}}d\theta ^{2}
\end{equation}%
in which%
\begin{multline}
f\left( \rho \right) =\tilde{C}_{1}\rho ^{\frac{2}{A^{2}+2}}-\frac{\lambda
_{1}\left( A^{2}+2\right) ^{2}}{2A^{2}\left( A^{2}-1\right) }\rho ^{\frac{%
2A^{2}}{A^{2}+2}}- \\
\frac{\left( \lambda _{2}+2q^{2}\right) \left( A^{2}+2\right) ^{2}}{2A^{2}}%
\rho ^{\frac{4}{A^{2}+2}}
\end{multline}%
for $\tilde{C}_{1}=\frac{C_{1}}{r_{0}^{2}}.$ We notice that the case with $%
A^{2}=2$ yields $q=0,$ $\alpha _{1}=\alpha _{2}=-2\sqrt{2}$ with the metric
function%
\begin{equation}
f\left( \rho \right) =\tilde{C}_{1}\sqrt{\rho }-4\left( \lambda _{1}+\lambda
_{2}\right) \rho
\end{equation}%
and the line element%
\begin{equation}
ds^{2}=-f\left( \rho \right) d\tau ^{2}+\frac{d\rho ^{2}}{f\left( \rho
\right) }+\rho d\theta ^{2}.
\end{equation}%
We set $\rho =x^{2}$ to get%
\begin{equation}
ds^{2}=-f\left( x\right) d\tau ^{2}+\frac{4x^{2}dx^{2}}{f\left( x\right) }%
+x^{2}d\theta ^{2},
\end{equation}%
with 
\begin{equation}
f\left( x\right) =\tilde{C}_{1}x-4\left( \lambda _{1}+\lambda _{2}\right)
x^{2}.
\end{equation}%
Depending on the sign of $\tilde{C}_{1}$ and $\lambda _{1}+\lambda _{2}$ the
corresponding spacetime can be black hole or not. For the case of the black
hole we consider $4\left( \lambda _{1}+\lambda _{2}\right) =-\frac{1}{\ell
^{2}}$ and $\tilde{C}_{1}=-a$ so that we find%
\begin{equation}
f\left( x\right) =\frac{x^{2}}{\ell ^{2}}-ax
\end{equation}%
with an event horizon at $x_{h}=a\ell ^{2}$ and line element%
\begin{equation}
ds^{2}=-\frac{x}{\ell ^{2}}\left( x-x_{h}\right) d\tau ^{2}+\frac{4x\ell
^{2}dx^{2}}{\left( x-x_{h}\right) }+x^{2}d\theta ^{2}.
\end{equation}%
The Ricci and Kretschmann scalars are found to be singular at $x=0,$ given
respectively by%
\begin{equation}
\mathcal{R}=-\frac{2x+x_{h}}{4x^{3}\ell ^{2}}
\end{equation}%
and%
\begin{equation}
\mathcal{K}=\frac{4x^{2}-4x_{h}x+3x_{h}^{2}}{16x^{6}\ell ^{4}}.
\end{equation}%
The scalar field reads%
\begin{equation}
\phi =\frac{\ln x}{\sqrt{2}}
\end{equation}%
with the potential%
\begin{equation}
V\left( \phi \right) =\frac{\lambda _{1}+\lambda _{2}}{x^{2}}.
\end{equation}%
The corresponding Hawking temperature is found as%
\begin{equation}
T_{H}=\left( \frac{-g_{tt}^{\prime }}{4\pi \sqrt{-g_{tt}g_{xx}}}\right)
_{x=x_{h}}=\frac{1}{8\pi \ell ^{2}},
\end{equation}%
which is constant. This signals an isothermal Hawking process in analogy
with a linear dilaton black hole in $3+1-$dimensions. Using the standard
entropy i.e., 
\begin{equation}
S=\pi x_{h}
\end{equation}%
one finds that the specific heat capacity becomes%
\begin{equation}
C_{q}=T_{H}\left( \frac{\partial S}{\partial T_{H}}\right) _{q}=0.
\end{equation}

\subsection{Exact solution with $A=1.$}

As we mentioned before, the case $A=1$ is excluded in the previous section.
Here we give the solution separately when $A=1.$ The field equations admit%
\begin{equation}
\phi =\frac{\ln r}{3}
\end{equation}%
and%
\begin{equation}
U\left( r\right) =C_{1}r^{2/3}-9q^{2}r^{4/3}-\frac{9\lambda _{2}r^{4/3}}{2}%
-3\lambda _{1}r^{2/3}\ln r
\end{equation}%
where $C_{1}$ is an integration constant. The line element may be written as%
\begin{equation}
ds^{2}=-U\left( y\right) dt^{2}+\frac{36y^{10}dy^{2}}{U\left( y\right) }%
+y^{2}d\theta ^{2},
\end{equation}%
in which we set $r^{1/3}=y^{2}$ and%
\begin{equation}
U\left( y\right) =y^{4}\left( C_{1}-9\left( q^{2}+\frac{\lambda _{2}}{2}%
\right) y^{4}-18\lambda _{1}\ln y\right) .
\end{equation}%
We note that the line element (36) can be a black hole or a naked singular
spacetime. Its Ricci scalar is given by%
\begin{equation}
\mathcal{R}=\frac{-10\lambda _{1}\ln y+11\left( q^{2}+\frac{\lambda _{2}}{2}%
\right) y^{4}+\frac{5C_{1}}{9}+4\lambda _{1}}{2y^{8}}.
\end{equation}%
In the case of the black hole with an event horizon at $y=y_{h}\neq 0$ one
finds the Hawking temperature%
\begin{equation}
T_{H}=-\frac{3\left( \left( 2q^{2}+\lambda _{2}\right) y_{h}^{4}+\lambda
_{1}\right) }{4\pi y_{h}^{2}}
\end{equation}%
and heat capacity as%
\begin{equation}
C_{q}=\frac{\pi y_{h}\left[ y_{h}^{4}\left( 2q^{2}+\lambda _{2}\right)
+\lambda _{1}\right] }{2\left[ y_{h}^{4}\left( 2q^{2}+\lambda _{2}\right)
-\lambda _{1}\right] }.
\end{equation}%
A phase change at $y_{h}^{4}\left( 2q^{2}+\lambda _{2}\right) -\lambda
_{1}=0 $ can occur but for the large enough horizon $C_{q}>0$ which
indicates the thermodynamical stability of the solution.

\section{Construction of the General solution}

Next, let us introduce $K=-2\phi ^{\prime 2},$ $R=\exp \left( \int
Ydr\right) $ and $u=UR$ which transform the field equations into the Riccati
form%
\begin{equation}
Y^{2}+Y^{\prime }=K,
\end{equation}%
with%
\begin{equation}
u^{\prime \prime }=-3V\exp \left( \int Ydr\right) -2q^{2}\exp \left( -\int
Ydr\right) ,
\end{equation}%
\begin{equation}
\left( u\phi ^{\prime }\right) ^{\prime }=\frac{\exp \left( \int Ydr\right) 
}{4}\frac{dV}{d\phi }
\end{equation}%
and%
\begin{equation}
\left( uY\right) ^{\prime }=-2q^{2}\exp \left( -\int Ydr\right) -V\exp
\left( \int Ydr\right) .
\end{equation}%
We combine (41) and (43) to eliminate $V$ i.e., 
\begin{equation}
u^{\prime \prime }-3\left( uY\right) ^{\prime }=4q^{2}\exp \left( -\int
Ydr\right) .
\end{equation}%
An integration implies%
\begin{equation}
u^{\prime }-3uY=4q^{2}\int dr\exp \left( -\int Ydr\right) +c_{1}
\end{equation}%
in which $c_{1}$ is an integration constant. Upon using $Y=\frac{R^{\prime }%
}{R}$ one finds%
\begin{equation}
u=R^{3}\left( \int \frac{1}{R^{3}}\left[ 4q^{2}\int \frac{dr}{R}+c_{1}\right]
dr+c_{2}\right)
\end{equation}%
in which $c_{2}$ is another integration constant. Having $u$ one finds from
(41)%
\begin{equation}
V=-\frac{u^{\prime \prime }}{3R}-\frac{2q^{2}}{3R^{2}}.
\end{equation}

\subsection{An Explicit Example}

To find an explicit solution one has to choose an ansatz for $\phi $ and
then by following the results given above, to find the other unknown
functions.

\subsubsection{$\protect\phi =\protect\alpha \ln \left( \frac{r}{r_{0}}%
\right) $}

Our choice for $\phi $ is a logarithmic function of the form 
\begin{equation}
\phi =\alpha \ln \left( \frac{r}{r_{0}}\right)
\end{equation}%
in which $\alpha $ and $r_{0}$ are two constants. Upon (48), the Riccati
equation becomes%
\begin{equation}
Y^{2}+Y^{\prime }=-2\phi ^{\prime 2}
\end{equation}%
with a solution of the form 
\begin{equation}
Y=\frac{A}{r}
\end{equation}%
in which $A$ is a constant, satisfying 
\begin{equation}
A^{2}-A+2\alpha ^{2}=0.
\end{equation}%
This condition imposes that $0<A<1.$ Consequently one finds 
\begin{equation}
R=\exp \left( \int Ydr\right) =\left( \frac{r}{r_{1}}\right) ^{A}
\end{equation}%
in which $r_{1}$ is an integration constant. Next, we find 
\begin{multline}
u=\frac{2q^{2}r_{1}^{2}}{\left( 1-A\right) \left( 1-2A\right) }\left( \frac{%
r_{1}}{r}\right) ^{A-2}+ \\
\frac{c_{1}}{1-3A}r+c_{2}\left( \frac{r}{r_{1}}\right) ^{3A}
\end{multline}%
for $A\neq 1/3,1/2.$ Finally, the potential $V$ is found to be%
\begin{multline}
V=\frac{2q^{2}\left( 1-A\right) }{3\left( 1-2A\right) }\left( \frac{r}{r_{1}}%
\right) ^{-2A}- \\
\frac{c_{2}A\left( 3A-1\right) }{r_{1}^{2}}\left( \frac{r}{r_{1}}\right)
^{2\left( A-1\right) }.
\end{multline}%
One can use the inverse transformation to find 
\begin{equation}
U=\frac{2q^{2}r_{1}^{2}}{\left( 1-A\right) \left( 1-2A\right) }R^{\frac{2}{A}%
-2}+\frac{c_{1}}{1-3A}r_{1}R^{\frac{1}{A}-1}+c_{2}R^{2}
\end{equation}%
and as a result%
\begin{multline}
V\left( \phi \right) =\frac{2q^{2}\left( 1-A\right) }{3\left( 1-2A\right) }%
\left( \frac{r_{1}}{r_{0}}\right) ^{2A}e^{-2A\phi /\alpha }- \\
\frac{c_{2}A\left( 3A-1\right) }{r_{1}^{2}}\left( \frac{r_{1}}{r_{0}}\right)
^{2\left( 1-A\right) }e^{2\left( 1-A\right) \phi /\alpha }.
\end{multline}

We note that, $\alpha ,$ $A,$ $r_{0},$ $r_{1},$ $c_{1}$ and $c_{2}$ are
parameters to be chosen provided the constraint (51) is satisfied. One of
the simplest choice of the parameters can be given if we set $q=c_{1}=0$ and 
$c_{2}=1.$ The line element, hence, becomes%
\begin{equation}
ds^{2}=U\left( -dt^{2}+d\theta ^{2}\right) +\frac{dr^{2}}{U}
\end{equation}%
in which $U=\left( \frac{r}{r_{1}}\right) ^{2A}.$ The potential,
accordingly, reads as%
\begin{equation}
V\left( \phi \right) =\frac{A\left( 3A-1\right) }{r_{1}^{2}}\left( \frac{%
r_{1}}{r_{0}}\right) ^{2\left( 1-A\right) }e^{2\left( 1-A\right) \phi
/\alpha }
\end{equation}%
in which $\phi $ is given by (48). It is observed that the case $A=1$ is not
included in our solution. The latter case has been found in \cite{7,8,9}.
Here in our solution $0<A<1$ and one of the simplest choice of $A$ is $\frac{%
1}{2}$ which yields $\alpha =\pm \frac{1}{2\sqrt{2}}$, $U=\frac{r}{r_{1}}$
and 
\begin{equation}
V\left( \phi \right) =\frac{1}{4r_{0}r_{1}}e^{\pm 2\sqrt{2}\phi }
\end{equation}%
while $\phi =$ $\pm \frac{1}{2\sqrt{2}}\ln \left( \frac{r}{r_{0}}\right) .$
(We note that with $q=c_{1}=0$ the solution for $A=\frac{1}{2}$ (for $q=0$)
is the same as (29)). The resulting line element, then, becomes (let's set $%
r_{0}=1$)%
\begin{equation}
ds^{2}=\left( \frac{r}{r_{1}}\right) \left( -dt^{2}+d\theta ^{2}\right)
+\left( \frac{r_{1}}{r}\right) dr^{2}.
\end{equation}%
Schmidt and Singleton found this solution in their work \cite{10} where $%
U=\left( \frac{r}{\ell }\right) ^{2},$ $\phi \sim \ln r$ and $V\left( \phi
\right) \sim e^{-2\sqrt{\kappa }\phi }.$

Our new solution which we shall proceed, begins with the case when $c_{1}=0,$
$c_{2}\neq 0$ and $A=\frac{2}{3}.$ The metric function then becomes%
\begin{equation}
U=-18q^{2}r_{1}^{2}R+c_{2}R^{2}
\end{equation}%
for 
\begin{equation}
R=\left( \frac{r}{r_{1}}\right) ^{\frac{2}{3}}.
\end{equation}%
This solution is not a black hole since a regular horizon doesn't exist. The
metric function becomes by the choice $c_{2}=18q^{2}r_{1}^{2}$%
\begin{equation}
U=\chi r^{\frac{2}{3}}\left( r^{\frac{2}{3}}-r_{1}^{\frac{2}{3}}\right)
\end{equation}%
in which $\chi =18q^{2}r_{1}^{\frac{2}{3}}.$

\subsection{A bounded regular solution}

In this section we set $q=0,$ $\phi =\frac{r}{r_{0}}$ in which for our later
convenience we set $r_{0}=\sqrt{2}.$ The Eq. (49) becomes%
\begin{equation}
Y^{2}+Y^{\prime }=-1
\end{equation}%
with the solution given by%
\begin{equation}
Y=-\tan r
\end{equation}%
and consequently%
\begin{equation}
R\left( r\right) =\left\vert \cos r\right\vert
\end{equation}%
and%
\begin{equation}
u\left( r\right) =\frac{\xi _{1}}{2}\cos r\left( \sin r+\cos ^{2}r\ln
\left\vert \sec r+\tan r\right\vert \right) +\xi _{2}\cos ^{3}r.
\end{equation}%
Finally%
\begin{equation}
U\left( r\right) =\frac{\xi _{1}}{2}\left( \sin r+\cos ^{2}r\ln \left\vert
\sec r+\tan r\right\vert \right) +\xi _{2}\cos ^{2}r
\end{equation}%
in which $\xi _{1}$ and $\xi _{2}$ are two integration constants and 
\begin{multline}
V\left( r\right) =\xi _{2}\left( -2+3\cos ^{2}r\right) + \\
\xi _{1}\left[ \frac{3}{2}\sin r+\frac{\left( -2+3\cos ^{2}r\right) }{2}\ln
\left( \frac{1+\sin r}{\cos r}\right) \right] .
\end{multline}%
With $\xi _{1}=0$ the line element becomes%
\begin{equation}
ds^{2}=-\xi _{2}\cos ^{2}rdt^{2}+\frac{dr^{2}}{\xi _{2}\cos ^{2}r}+\cos
^{2}rd\theta ^{2}.
\end{equation}%
Next, we apply the following change of variable%
\begin{equation}
x=\cos r
\end{equation}%
which implies%
\begin{equation}
ds^{2}=-\xi _{2}x^{2}dt^{2}+\frac{dx^{2}}{\xi _{2}x^{2}\left( 1-x^{2}\right) 
}+x^{2}d\theta ^{2}
\end{equation}%
with $\left\vert x\right\vert <1$ whose scalar curvature and Kretschmann
scalar are 
\begin{equation}
\mathcal{R}=\xi _{2}\left( 10x^{2}-6\right)
\end{equation}%
and%
\begin{equation}
\mathcal{K}=4\xi _{2}^{2}\left( 17x^{4}-20x^{2}+6\right) .
\end{equation}%
The potential, becomes%
\begin{equation}
V\left( \phi \right) =\xi _{2}\left( -2+3\cos ^{2}\left( \sqrt{2}\phi
\right) \right)
\end{equation}%
with the scalar field%
\begin{equation}
\phi =\frac{\arccos x}{\sqrt{2}}.
\end{equation}%
We apply now a new transformation which maps $x\in \left( -1,1\right) $ into 
$\rho \in \left[ 0,\infty \right) $ as defined by%
\begin{equation}
\rho =\sqrt{\frac{x^{2}}{1-x^{2}}}
\end{equation}%
with the transformed line element (we note that $\xi _{2}>0$ has no role in
the scalars and therefore we set it as $\xi _{2}=1$) 
\begin{equation}
ds^{2}=\frac{1}{1+\rho ^{2}}\left( -\rho ^{2}dt^{2}+\frac{d\rho ^{2}}{\rho
^{2}}+\rho ^{2}d\theta ^{2}\right) .
\end{equation}%
Also the scalar field becomes%
\begin{equation}
\phi =\frac{\arccos \left( \frac{\rho }{\sqrt{1+\rho ^{2}}}\right) }{\sqrt{2}%
}
\end{equation}%
%
%
\begin{figure}[h]
\includegraphics[width=80mm,scale=0.7]{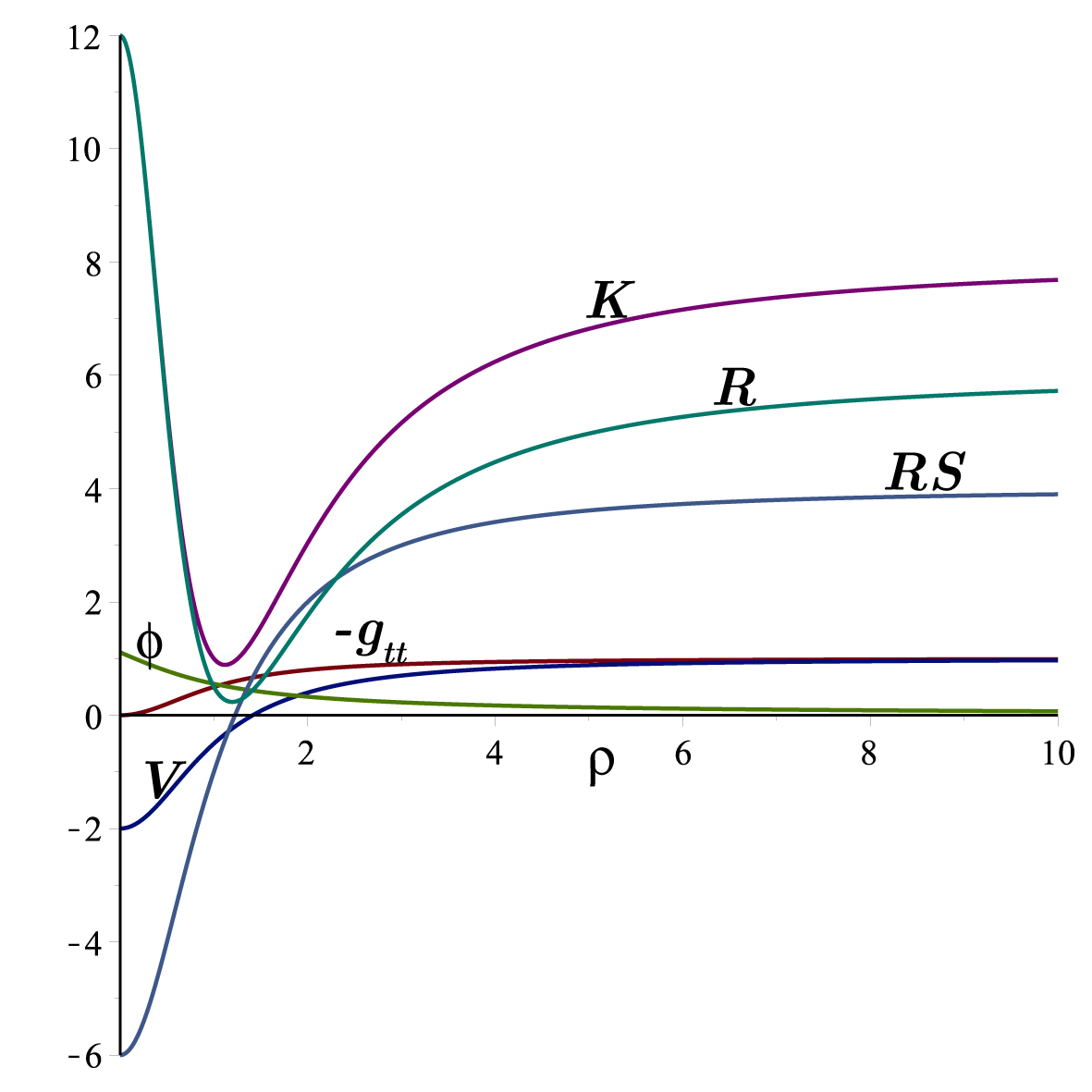}
\caption{A plot of $V\left( \protect\rho \right) ,$ $\protect\phi \left( 
\protect\rho \right) ,$ $\mathcal{R}$, $\mathcal{RS}$, $\mathcal{K}$ and $%
-g_{tt}$ with respect to $\protect\rho .$ The astrophysical object is
regular everywhere and asymptotically, for $\protect\rho \rightarrow \infty
, $ it becomes conformal to the AdS spacetime..}
\end{figure}
with the potential 
\begin{equation}
V\left( \rho \right) =\frac{\rho ^{2}-2}{\rho ^{2}+1}.
\end{equation}%
It is observed that the scalars take the forms%
\begin{equation}
\mathcal{R}=\frac{2\left( 2\rho ^{2}-3\right) }{1+\rho ^{2}}
\end{equation}%
\begin{equation*}
\mathcal{RS}=R_{\mu \nu }R^{\mu \nu }=\frac{2\left( 6-8\rho ^{2}+3\rho
^{4}\right) }{\left( 1+\rho ^{2}\right) ^{2}}
\end{equation*}%
and%
\begin{equation}
\mathcal{K}=\frac{4\left( 3-4\rho ^{2}+2\rho ^{4}\right) }{\left( 1+\rho
^{2}\right) ^{2}}.
\end{equation}%
The spacetime given by (78) is not a black hole and also is not singular.
This is a regular object with scalar invariant everywhere in the domain of $%
\rho \in \left[ 0,\infty \right) .$ In Fig. 1 we plot $V\left( \rho \right) $%
, $\phi \left( \rho \right) $, $\mathcal{R}$, $\mathcal{RS}$, $\mathcal{K}$
and $-g_{tt}$ in terms of $\rho $ which display the general behavior of the
cosmological object in the domain of $\rho .$

\section{Conclusion}

To what extent self-interacting scalar fields play role in lower dimensions
such as $2+1$?. Can scalar charge be chosen to imitate the role of electric
charge in making of black holes?. These are the problems that we addressed /
answered in this paper. We obtained both exact black hole and non-black hole
solutions described by potentials of the form $V\left( \phi \right) \sim
\lambda _{1}e^{\alpha _{1}\phi }+\lambda _{2}e^{\alpha _{2}\phi }$ with $%
\lambda _{i}$ and $\alpha _{i}$ constants, coupled to a charged mass / black
hole. In particular, we obtain black hole solutions with zero electric
charge when the parameters are tuned. The non-black hole solutions give rise
to singularities which are strongly naked. The system is described
effectively by a Riccati type differential equation. By changing the ansatz
for the scalar field our model can be shown to admit different classes of
solutions so that the solution space is quite large. As a particular example
we present a bounded, completely regular solution which is asymptotically,
i.e. for $\rho \rightarrow \infty ,$ conformal to the AdS spacetime. This
admits a finite potential-well at the origin to attract interest from field
theoretical point of view.

\end{document}